\documentclass{iau}
\usepackage{graphicx}

\title[Large-scale peculiar velocities through the galaxy luminosity function] 
{Large-scale peculiar velocities through the galaxy luminosity function at \boldmath{$z\sim 0.1$}}

\author[Feix, Nusser \& Branchini]   
{Martin Feix$^{1}$, Adi Nusser$^{1,2}$, \and Enzo Branchini$^{3,4,5}$}

\affiliation{$^1$Department of Physics, Israel Institute of Technology - Technion, Haifa 32000, Israel \\
email: {\tt mfeix@physics.technion.ac.il} \\[\affilskip]
$^2$Asher Space Science Institute, Israel Institute of Technology - Technion, Haifa 32000, Israel \\[\affilskip]
$^3$Department of Physics, Universit\`a Roma Tre, Via della Vasca Navale 84, Rome 00146, Italy \\[\affilskip]
$^4$INFN Sezione di Roma 3, Via della Vasca Navale 84, Rome 00146, Italy \\[\affilskip]
$^5$INAF, Osservatorio Astronomico di Roma, Monte Porzio Catone, Italy}

\pubyear{2014}
\volume{308}  
\jname{The Zeldovich Universe: Genesis and Growth of the Cosmic Web}
\editors{R. van de Weygaert, S. Shandarin, E. Saar \& J. Einasto, eds.}
\begin{document}

\maketitle

\begin{abstract}
Peculiar motion introduces systematic variations in the observed luminosity distribution of galaxies. This allows one to constrain the
cosmic peculiar velocity field from large galaxy redshift surveys. Using around half a million galaxies from the SDSS Data Release $7$
at $z\sim 0.1$, we demonstrate the applicability of this approach to large datasets and obtain bounds on peculiar velocity moments and
$\sigma_{8}$, the amplitude of the linear matter power spectrum. Our results are in good agreement with the $\Lambda$CDM model and
consistent with the previously reported $\sim1$\% zero-point tilt in the SDSS photometry. Finally, we discuss the prospects of
constraining the growth rate of density perturbations by reconstructing the full linear velocity field from the observed galaxy
clustering in redshift space.

\end{abstract}

\firstsection 
\section{Velocities from the variation of observed galaxy luminosities}
To linear order in perturbation theory, the observed redshift $z$ of a galaxy typically deviates from its cosmological redshift $z_{c}$
according to (\cite[Sachs \& Wolfe 1967]{SW1967})
\begin{displaymath}
\frac{z-z_{c}}{1+z} = \frac{V(t,r)}{c} - \frac{\Phi(t,r)}{c^2} - \frac{2}{c^2}\int_{t(r)}^{t_0}{\rm d}t
\frac{\partial\Phi\left\lbrack\hat{\boldsymbol{r}}r(t),t\right\rbrack}{\partial t}\approx \frac{V(t,r)}{c},
\end{displaymath}
where $V$ is the (physical) radial peculiar velocity of the galaxy, $r$ is a unit vector along the line of sight to the object, and
$\Phi$ denotes the usual gravitational potential. Here we explicitly assume low redshifts such that the velocity $V$ is the dominant
contribution, and we further consider all fields relative to their present-day values at $t_{0}$.

As the shift $z-z_{c}$ enters the calculation of distance moduli ${\rm DM}=25+5\log_{\rm 10}\lbrack D_{L}/{\rm Mpc}\rbrack$, where
$D_{L}$ is the luminosity distance, observed absolute magnitudes $M$ differ from their true values $M^{(t)}$. We thus have
\begin{displaymath}
M = m - {\rm DM}(z) - K(z) + Q(z) = M^{(t)} + 5\log_{10}\frac{D_{L}(z_{c})}{D_{L}(z)},
\end{displaymath}
where $m$ is the apparent magnitude, the function $Q(z)$ accounts for luminosity evolution, and $K(z)$ is the $K$-correction
(\cite[Blanton \& Roweis 2007]{Blanton2007}). On scales where linear theory provides an adequate description, the variation $M-M^{(t)}$
of magnitudes distributed over the sky is systematic, and therefore, contains information on the peculiar velocity field.

Given a suitable parameterized model $V(\hat{\boldsymbol{r}}, z)$ of the radial velocity field, the idea is now to maximize the
probability of observing galaxies with magnitudes $M_{i}$ given only their redshifts and angular positions $\hat{\boldsymbol{r}}_{i}$
on the sky, i.e.,
\begin{displaymath}
P_{\rm tot} = \prod\limits_{i}P\left (M_{i}\vert z_{i}, V(\hat{\boldsymbol{r}}_{i},z_{i})\right ) =
\prod\limits_{i}\left (\phi(M_{i})\middle /\int_{M_{i}^{+}}^{M_{i}^{-}}\phi(M){\rm d}M\right ),
\end{displaymath}
where we assume that redshift errors can be neglected (\cite[Nusser \etal\ 2011]{Nusser2011}), $\phi(M)$ denotes the
galaxy luminosity function (LF), and the corresponding limiting magnitudes $M^{\pm}$ depend on $V(\hat{\boldsymbol{r}}, z)$ through the
cosmological redshift $z_{c}$. Here the motivation is to obtain a maximum-likelihood estimate of $V(\hat{\boldsymbol{r}}, z)$ by finding
the set of velocity model parameters which minimizes the spread in the observed magnitudes.

\cite[Tammann \etal\ (1979)]{Tammann1979} first adopted this approach to estimate the motion of Virgo relative to the local group, and
recently, \cite[Nusser \etal\ (2011)]{Nusser2011} used it to constrain bulk flows in the local Universe from the 2MASS Redshift Survey
(\cite[Huchra \etal\ 2012]{Huchra 2012}).

\section{Constraints on the cosmic peculiar velocity field at \boldmath{$z\sim 0.1$}}
Galaxies from the Sloan Digital Sky Survey (SDSS) Data Release 7 (\cite[Abazajian \etal\ 2009]{Abazajian2009}) probe the cosmic velocity
field out to $z\sim 0.1$. Here we report results obtained from applying the luminosity method to a subset of roughly half a million
galaxies (\cite[for additional details, see Feix \etal\ 2014]{Feix2014}).

{\underline{\it Data}}. In our analysis, we used the latest version of the NYU Value-Added Galaxy Catalog (NYU-VAGC; \cite[Blanton \etal\
2005]{Blanton2005}). Giving the largest spectroscopically complete galaxy sample, we adopted (Petrosian) $^{0.1}r$-band magnitudes, and
chose the subsample NYU-VAGC {\tt safe} to minimize incompleteness and systematics. Our final sample contained only galaxies with $14.5 <
m_{r} < 17.6$, $-22.5 < M_{r} - 5\log_{10}h < -17.0$, and $0.02 < z < 0.22$ (relative to the CMB frame). In addition, we employed a suite
of galaxy mock catalogs mimicking the known systematics of the data.

{\underline{\it Radial velocity model}}. We considered a bin-averaged velocity model $\tilde{V}(\hat{\boldsymbol{r}})$ in two redshift
bins, $0.02 < z < 0.07$ and $0.07 < z < 0.22$. For each bin, the velocity field was further decomposed into spherical harmonics, i.e.
\begin{displaymath}
a_{lm} = \int{\rm d}\Omega\tilde{V}(\hat{\boldsymbol{r}})Y_{lm}(\hat{\boldsymbol{r}}),\qquad \tilde{V}(\hat{\boldsymbol{r}})
= \sum\limits_{l,m}a_{lm}Y_{lm}^{*}(\hat{\boldsymbol{r}}),\qquad l>0,
\end{displaymath}
where the sum over $l$ is cut at some maximum value $l_{\rm max}$. Because the SDSS data cover only part of the sky, the inferred $a_{lm}$
are not statistically independent. The impact of the angular mask was studied with the help of suitable galaxy mock catalogs. The monopole
term ($l=0$) was not included since it is degenerate with an overall shift of magnitudes.

{\underline{\it LF estimators}}. Reliably measuring the galaxy LF represents a key step in our approach. To assess the robustness of our
results with respect to different LF models, we analyzed the data using LF estimators based on a Schechter form and a more flexible
spline-based model, together with several combinations and variations thereof. For simplicity, we also assumed a linear dependence of the
luminosity evolution with redshift.

{\underline{\it Bulk flows and higher-order velocity moments}}.
Accounting for known systematic errors in the SDSS photometry, our ``bulk flow'' measurements are consistent with a standard $\Lambda$CDM
cosmology at a $1$--$2\sigma$ confidence level in both redshift bins. A joint analysis of the corresponding three Cartesian components
confirmed this result. To characterize higher-order moments as well, we further obtained direct constraints on the angular velocity power
spectrum $C_{l} = \langle |a_{lm}|^{2}\rangle$ up to the octupole contribution. The estimated $C_{l}$ were found compatible to be with the
theoretical power spectra of the $\Lambda$CDM cosmology.

\begin{figure}[t]
\begin{center}
 \includegraphics[width=0.95\linewidth]{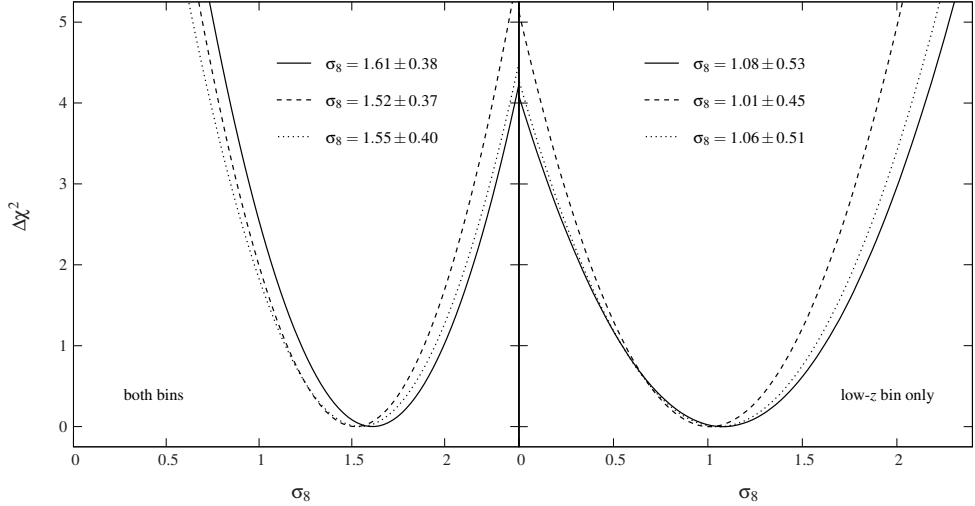} 
 \caption{Raw estimates of $\sigma_{8}$ obtained from the NYU-VAGC: shown is the derived $\Delta\chi^{2}$ as a
 function of $\sigma_{8}$ for both redshift bins (left panel) and the first redshift bin with $0.02<z<0.07$ only
 (right panel), adopting different estimators of the LF (solid, dashed, and dotted lines).}
   \label{fig1}
\end{center}
\end{figure}

\begin{figure}[b]
\begin{center}
 \includegraphics[width=0.95\linewidth]{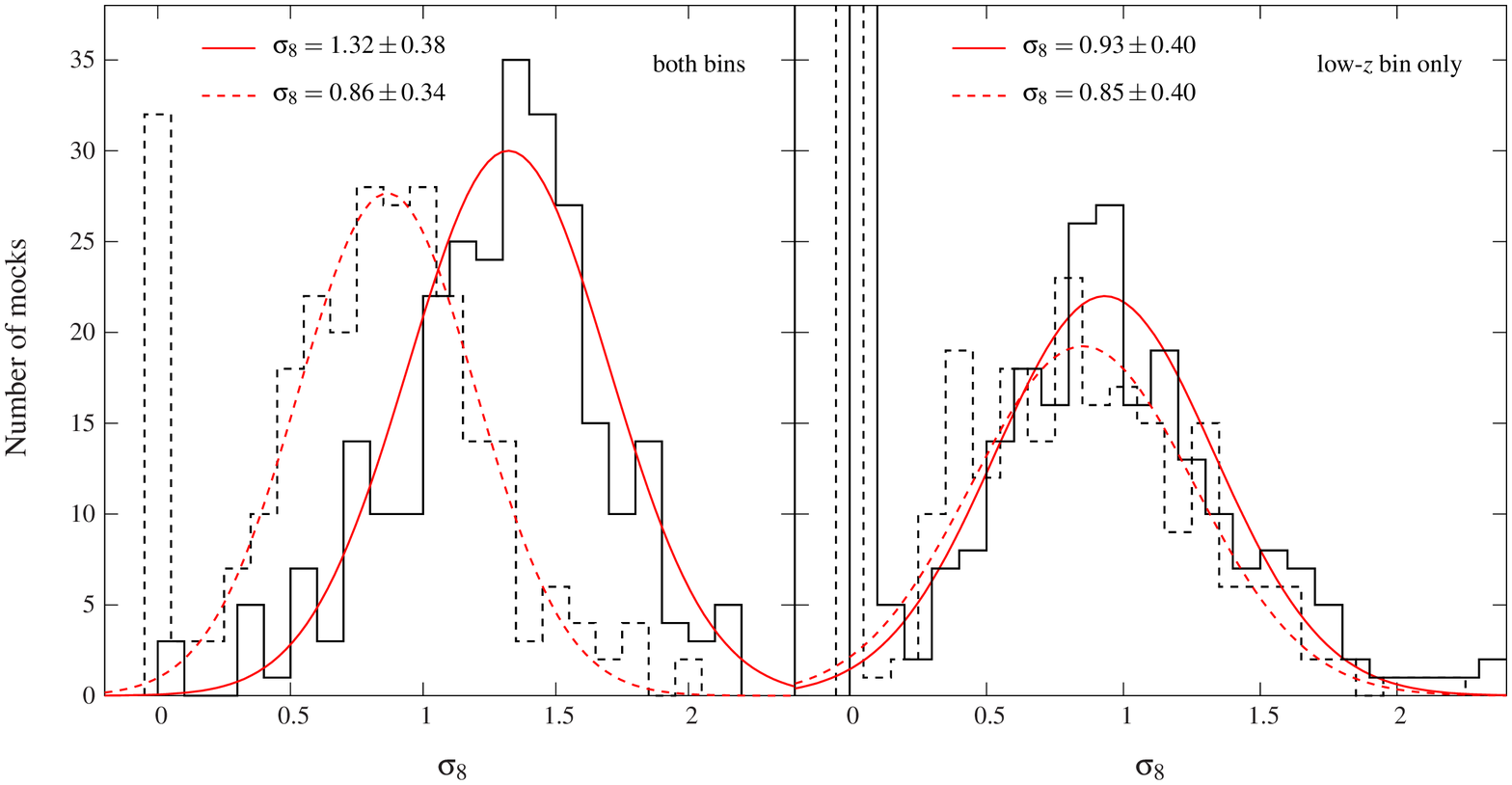} 
 \caption{Distribution of $\sigma_{8}$ estimated from mock galaxy catalogs: shown are the recovered histograms (black lines) and
 respective Gaussian fits with (solid lines) and without (dashed lines) the inclusion of a systematic (randomly oriented) tilt in
 the galaxy magnitudes, using the information in both redshift bins (left) and the bin with $0.02<z<0.07$ only (right).}
   \label{fig2}
\end{center}
\end{figure}

{\underline{\it Constraints on $\sigma_{8}$}}.
Assuming a prior on the $C_{l}$ as dictated by the $\Lambda$CDM model with fixed Hubble constant and density parameters, we
independently estimated the parameter $\sigma_{8}$ which determines the amplitude of the velocity field. Due to the presence of a
dipole-like tilt in the galaxy magnitudes (\cite[Padmanabhan \etal\ 2008]{Padmanabhan2008}), the obtained raw estimates of $\sigma_{8}$
were expected to be biased toward larger values (Fig.\,\ref{fig1}). After correcting for this magnitude tilt with the help of our
mocks (Fig.\,\ref{fig2}), we eventually found $\sigma_{8}\approx 1.1\pm 0.4$ for the combination of both redshift bins and $\sigma_{8}
\approx 1.0\pm 0.5$ for the low-$z$ bin only, where the low accuracy is due to the limited number of galaxies. This confirms our
method's validity in view of future datasets with larger sky coverage and better photometric calibration.

\section{Toward constraints on the linear growth rate}
A very interesting aspect of our luminosity-based approach is the possibility to place bounds on the growth rate of density perturbations,
$\beta = f(\Omega)/b$ (where $b$ is the linear galaxy bias), by modeling the large-scale velocity field directly from the observed
clustering of galaxies in redshift space (\cite[Nusser \& Davis 1994]{Nusser1994}). Such bounds are complementary to and --- regarding
ongoing and future redshift surveys --- expected to be competitive with those obtained from redshift-space distortions
(\cite[Nusser \etal\ 2012]{Nusser2012}).

To get an idea of how well the method could constrain $\beta$ at $z\sim 0.1$ from SDSS galaxies, we used mocks generated from the
Millennium Simulation (\cite[Springel \etal\ 2005]{Springel2005}; \cite[Henriques \etal\ 2012] {Henriques2012}) to create full-sky
catalogs which otherwise shared all characteristics of the real SDSS data. Adopting a radial velocity model proportional to the true one
smoothed over spheres of $10h^{-1}$ Mpc radius, the luminosity method was applied to samples with around $2\times 10^{5}$ galaxies and
correctly recovered the velocity field. The error on the proportionality constant typically yielded $\pm 0.2$--$0.3$ if only the
contribution of multipoles with $l > 25$ (an appropriate value for the SDSS geometry) is taken into account. Assuming an accurate
velocity reconstruction for these modes, we expect a similar situation for $\beta$. A further complication is that the angular mask
may introduce bias as a consequence of multipole mixing. This and other technical issues mainly related to the reconstruction of the
velocity field are currently under detailed investigation.

\section{Outlook}
Current and next-generation spectroscopic surveys are designed to reduce data-inherent systematics because of larger sky coverage
and improved photometric calibration in ground- and space-based experiments (e.g, \cite[Levi \etal\ 2013]{Levi2013}; \cite[Laureijs \etal\
2011]{Laureijs2011}). The method considered here does not require accurate redshifts and can be used with photometric redshift surveys
such as the 2MASS Photometric Redshift catalog (2MPZ; \cite[Bilicki]{Bilicki2014} \cite[\etal\ 2014]{Bilicki2014}) to recover signals on
scales larger than the spread of the redshift error.

Together with our results, these observational perspectives give us confidence that the luminosity-based method will be established
as a standard cosmological probe, independent from and alternative to the more traditional ones based on galaxy clustering, gravitational
lensing and redshift-space distortions.

\end{document}